\documentclass[12pt]{iopart}
\input epsf

\include{psfig}

\begin{document}
\title{Thermodynamics of supercooled liquids in the inherent structure 
formalism: a case study}

\author{Francesco Sciortino$^{(1)}$, Walter Kob$^{(2)}$, 
Piero Tartaglia$^{(1)}$}

\address{$^{(1)}$ Dipartimento di Fisica and Istituto Nazionale
 per la Fisica della Materia, Universit\'a di Roma {\it La Sapienza},\\
 P.le Aldo Moro 2, I-00185, Roma, Italy \\
$^{(2)}$   Institut f\"ur Physik, Johannes Gutenberg--Universit\"at,
Staudinger Weg 7, D-55099 Mainz, Germany}

\begin{abstract}
In this article we review the thermodynamics of liquids in the
framework of the inherent structure formalism. 
We then present calculations of the distribution of the basins in the 
potential energy of
a binary Lennard-Jones mixture as a function of temperature.
The comparison between the numerical data and the theoretical formalism
allows us to evaluate 
the degeneracy of the inherent structures in a bulk system and to estimate
the energy of the lowest energy disordered state (the Kauzmann energy).
We find that, around the mode-coupling temperature,
the partition function of the liquid is approximated well by the product of
two loosely coupled partition functions, one depending on the
inherent structures quantities (depth of the basins and their degeneracy) 
and one describing the free energy of the liquid constrained in one
typical basin. 

\end{abstract}


\submitted{{\noindent \it }}

\section{Introduction}

The potential energy of a system composed by $N$ interacting atoms is
a complicated surface in a $3N$-dimension space. The motion of the system
can be thought of as a trajectory over such a potential energy surface
(PES). At different temperatures, the system explores different parts
of the PES, according to the Boltzmann weight. The idea of focusing on the
PES for understanding the physics of glass forming liquids can 
be traced back to 
the seminal (but talkative) paper of Goldstein \cite{goldstein}.
He suggested that the dynamics of deeply supercooled
liquids can be described in terms of a diffusive process of the system
between different PES basins. At low temperatures the dynamics slows down 
since the liquid becomes trapped for a long time in a basin.

The concept of a basin in configuration space
was formalized by Stillinger and Weber\cite{sw}, who introduced 
a recipe, very well suited for numerical analysis, for
partitioning the PES into disjoint basins. 
The set of  points in configuration space 
connected to the same local minimum via a steepest-descent trajectory defines
uniquely
the basin associated with this local minimum. Stillinger and Weber named
the structure of the  system in the minimum  inherent-structure (IS) and the 
value of the PES at the minimum inherent-structure energy ($e_{IS}$).

The increased computational facilities have significantly improved the
early efforts of studying the PES. Nowadays,
an exhaustive search for all IS has been
performed for clusters 
and complete maps of the inherent structure energies are 
available for several potential models\cite{wales}. For clusters,
as well as for small proteins\cite{walesp}, 
the connectivity between all IS has also been evaluated, 
to provide a very informative map both of the thermodynamics\cite{doye}
 as well 
as of the dynamics in these small systems\cite{kunz}.
Small size systems, composed by 30 to 50 atoms with periodic boundary
conditions, have also been studied in detail recently and 
almost exhaustive enumerations of all IS energies are now 
available\cite{stillingersearxh,heuer,angelani}.

In this article we review 
a theoretical framework in which the IS results can be
interpreted in a convenient way (Sec. \ref{sec:theory}) 
and discuss the approximations requested for
a factorization of the partition function in two functions, one describing the 
thermodynamics of the IS sub-system and one describing the thermodynamic
of the exploration of one representative basin (Sec.\ref{sec:theorylowT})
In the following two sections we present calculations of the temperature
(Sec. \ref{sec:entropyT}) and IS energy (Sec. 
\ref{sec:entropyE}) dependence of the
configurational entropy for a bulk system. 
Such calculations allow to quantify the properties of the PES for
model systems and 
to probe the validity of
the factorization approximation. New information on the
equilibrium and aging dynamics of supercooled simple 
liquids is provided by the presented results.
An account of the results has been reported in Ref.\cite{skt}

\section{Theory}
\label{sec:theory}
This section reviews the thermodynamic formulation proposed by
Stillinger and Weber\cite{sw}, 
focusing on the concept of basins in configuration space.

The partition function of a system composed by $N$ identical 
atoms of mass $m$, after the integration over the momentum variables is
\begin{equation}
Z_{N}= \lambda^{-3N}
\int \exp(-\beta V({\bf r}^N)) d{\bf r}^N
\end{equation}
where $\lambda = h \sqrt{\beta /2 \pi m}$.
The integral over the configuration space ${\bf r}^N$ can be
separated into a sum over all distinct basins
\begin{equation}
Z_{N}=\lambda^{-3N} \sum_{\alpha} \exp(-\beta \Phi_{\alpha})
\int_{R_{\alpha}} \exp(-\beta \Delta_{\alpha} ({\bf r}^{N})) d{\bf r}^N
\end{equation}
where $R_\alpha$ is the set of points composing the basin $\alpha$, 
$\Phi_{\alpha}$ is the potential energy of the minimum $\alpha$ 
and the non-negative quantity $ \Delta_{\alpha} ({\bf r}^N)$  
measures the potential energy at a point ${\bf r}^N$  
belonging to the basin $\alpha$ relative to the minimum.
By classifying the minima according to their IS energy $e_{IS}$,
the sum over the basins can be separated in a sum over all possible values of
 $e_{IS}$ and a sum over all basins $\alpha'$ with the same $e_{IS}$
value.
\begin{equation}
Z_{N} = \lambda^{-3N} \sum_{e_{_{IS}}} 
\exp(-\beta e_{_{IS}})\sum_{\alpha'} 
\int_{_{_{R_{\alpha'}}}} \exp(-\beta \Delta_{\alpha'} ({\bf r}^{^{N}}))~
d{\bf r}^N 
\end{equation}
Following Stillinger and Weber, we introduce 
an IS density of states $ \Omega(e_{_{IS}})$, which 
counts 
the number of distinct basins with $IS$ energy between $ e_{_{IS}}$  and 
$e_{_{IS}}+\delta e_{_{IS}}$ and define a basin free energy
$f(\beta, e_{_{IS}})$ as the average $e_{IS}$ basin free energy, according to 

\begin{equation}
-\beta f(\beta, e_{IS})  \equiv 
\ln \left(
{  \lambda^{-3N} \sum_{_{\alpha'}} 
 \int_{_{R_{\alpha'}}} \exp(-\beta \Delta_{\alpha'} ({\bf r}^N))~
d{\bf r}^N}  \over { \delta e_{_{IS}}  \Omega(e_{_{IS}}) }
\right)
\end{equation}

\noindent
If all basins with the same $e_{IS}$ energy have the same
 statistical properties then $f(\beta, e_{_{IS}})$ coincides with the free energy of a system 
constrained to sample only one basin and which does not know of the existence
of the other equivalent $ \delta e_{_{IS}}  \Omega(e_{_{IS}})$ basins.
$Z_N$ can be expressed in terms of PES quantities, as
\begin{equation}
Z_{N} =
\int de_{_{IS}} \Omega(e_{_{IS}})
\exp(-\beta e_{IS}- \beta f(\beta, e_{IS}))
\label{eq:zis}
\end{equation}
Performing a maximum integral evaluation of the partition function, 
the free energy $F$ of the system
can be expressed in the thermodynamic limit as 

\begin{equation}
F=  e^*_{IS}- T S_{conf}(e^*_{IS}) +
 f(\beta, e^*_{IS})
\end{equation}

\noindent
where $e^*(T)$ is the $e_{IS}$ value which maximizes the integrand
and $S_{conf}(e_{IS}) = k_B \ln(\delta e_{IS} \Omega(e_{IS}))$.
If we separate now $f(\beta, e)$ in its energetic $u_b$ and entropic $s_b$ 
contribution, we immediately notice that the entropy associated with the
basin degeneracy, 
$ S_{conf}(T)$, can be calculated as the difference between 
the system entropy and
$s_b$, the entropy of the system {\it constrained} in an IS $e^*$.
In the present formalism, if one excludes from the sum in the
partition function the crystalline $IS$, then one can 
identify $F$ as the fluid free energy for all $T$.

The choice of separating the liquid free energy in a sum of two
inter-related contributions (via the $e_{IS}$ dependence of  $f$)
has been often used in the past, for example
in estimating the configurational entropy from available 
experimental data\cite{wallace}. 
In this case, the basin entropy is identified with the
entropy of the stable crystal at the same thermodynamics point. Such an
identification is based on the idea that the vibrational properties
of a system constrained in a deep basin are similar to the
properties of the close crystalline structure.
More recently, the consequences of such separation for several
thermodynamic quantities have been explicitly worked out\cite{pablo}.
For recent related work see also \cite{eosel,andreas}

\section{Low $T$ approximation}
\label{sec:theorylowT}

There are two interesting cases which may help understanding the low
$T$ dynamics of liquids. These cases are connected to specific forms
of $f(\beta,e_{IS})$.  The first describes the case where 
$f(\beta,e_{IS}) \approx
f(\beta)$, i.e. there is no $T$-dependence through $e_{IS}$.
In this case basins are characterized by
approximately the same shape in configuration space, an hypothesis
which can be tested by studying the $e_{IS}$ dependence of the density
of states. In this approximation Eq.~(\ref{eq:zis}) factorizes in
\begin{equation}
Z_N \approx Z^{IS} \cdot Z^{b}
\label{eq:fact}
\end{equation}
where
\begin{equation}
Z^{IS} = \int \Omega(e_{_{IS}}) \exp(-\beta e_{IS}) de_{IS} 
\end{equation}
and 
\begin{equation}
Z^{b} =  \exp( - \beta f(\beta))
\end{equation}
In the range of $T$ where this approximation holds, the system can be
tough of as two weakly coupled subsystem: the $IS$-subsystem, 
which has now been transformed in a system with levels labeled by the $e_{IS}$ 
value with degeneracy $\Omega(e_{IS})$, and the basin subsystem 
which describe the motion in the characteristic basin.
The coupling between the two subsystem, which of course allows for the
equilibration process between the two subsystems, is due to the weak
$T$ dependence of $e_{IS}$, which is neglected in the present approximation.

The second case is the case when $\beta f(\beta,e_{IS}) \approx
g(\beta) + h(e_{IS}) $, i.e. when the $T$ and $e_{IS}$ dependence 
are not mixed. This case is realized for example in the case
where at different $T$ the system populate basins which are always
{\it harmonic}, but different in their density of states\cite{andreas}. 
In this second case, 
a factorization of $Z_N$ as in Eq.~(\ref{eq:fact}) is also possible
by redefining the density of states 
to include the basin volume in configuration space, as
$\Omega(e_{_{IS}}) e^{h(e_{IS})}$\cite{andreas}.

Analysis of computer simulation data allows to look for the existence of a
$T$ range where the
factorization approximation holds. Indeed, the probability density of 
extracting from a system in 
thermal equilibrium at temperature $T$ a configuration belonging to a
basin with IS energy $e_{IS}$ is
 
\begin{equation}
P(e_{IS},T)= {{ \Omega(e_{_{IS}})
\exp(-\beta e_{IS}- \beta f(\beta, e_{IS})) }\over {Z_N(T)}}
\end{equation}

\noindent
If the factorization approximation holds, then the only $ e_{IS}$ dependence
in
the rhs of 
\begin{equation}
\ln[P(e_{_{IS}},T)\delta e_{_{IS}} ] + \beta e_{_{IS}} = S_{conf}(e_{IS}) /k_B
 - \ln[ Z_N(\beta)] -  \beta f(\beta,e_{IS}) 
\label{eq:pofe}
\end{equation}

\noindent 
is contained in $S_{conf}$. This imply that
curves 
for different $T$ can be superimposed after a 
shift of a $T$-dependent quantity. The resulting $e_{IS}$
master curve is, except for an unknown constant,
the $e_{IS}$ configurational entropy. 

\section{System}
We have studied the 
well-known 80-20 Lennard-Jones
$A-B$ binary mixture (BMLJ), 
composed by 1000 atoms in a volume $V_o=(9.4)^3$,
corresponding to a reduced density of 1.2039. Units of length,
energy and are defined by the $\sigma$ and $\epsilon$ 
of the $A-A$  Lennard Jones interaction potential, and the 
unit of mass by the mass of atom $A$. The pair potential is defined in Ref.\cite{kob}.
The equilibrium and out-of-equilibrium slow dynamics has been
studied extensively. The
critical temperature of Mode Coupling Theory for this system is
$0.435$\cite{kob}.

New simulations, covering the range $0.446 \leq T \leq 5.0 $, 
have been performed in 
the canonical ensemble by coupling the system to 
a Nose'-Hoover thermostat\cite{nh}. 
~From simulations lasting more than 60M steps, we have extracted 1000
equally spaced configurations and we have calculated for 
each of them the corresponding IS. 

\section{Temperature Dependence of the configurational entropy}
\label{sec:entropyT}

We have estimated the  $T$-dependence configurational entropy for the 
BMLJ as difference of the liquid entropy and of the basin entropy, 
as discussed in section \ref{sec:theory}. 
An independent (and consistent) estimate
of the same quantity in the same system has been performed 
with a similar procedure by Coluzzi et al. \cite{btesi,barbara}.  
For related work see also Refs.~\cite{TKreferences}

The entropy of the liquid has been calculated via thermodynamic integration
starting from the ideal gas binary mixture reference point, 
along the $T=5.0$ isotherm, up to the studied $\rho=1.2$ density.
In the following we call $C$ the state point $T=5.0,\rho=1.2$. 
The entropy of the liquid in $C$
can be written as

\begin{equation}
S(C) =
S_{ideal-gas}(C)   
+ {{U(C)}\over {T}} +
\int_\infty ^{V_o} {{P_{ex} dV} \over {T}}
\label{eq:entro}
\end{equation}

\noindent
where
\begin{equation}
{{S_{ideal-gas}(T,\rho)} \over {N k_B}} =
 -{N_A \over N} \ln ({N_A \over N} )
- {N_B \over N} \ln ({N_B \over N} ) 
+ {3 \over 2} \ln ({ {e m V^{{2}\over{3}} }  \over { \beta \hbar^2 2 \pi }})
-\ln ( {{ N}\over {e}})
\end{equation}

\begin{figure}
\centerline{\psfig{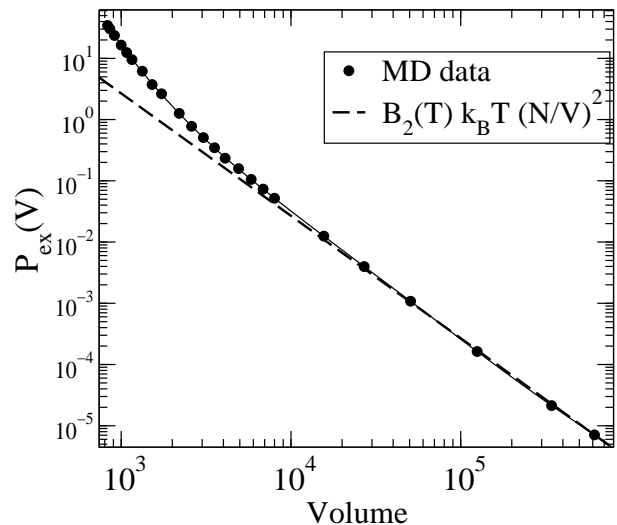}}
\caption{Excess pressure as a function of the
volume at $T=5.0$. The dashed line is the analytically calculated first virial
correction to the pressure.}
\label{fig:pex}
\end{figure}

\noindent
$P_{ex}$ is the excess pressure over the ideal gas value, $U$ is
the potential energy, and $e$ is the energy per particle. 
Fig. \ref{fig:pex} shows the excess pressure as a function of the
volume calculated from twenty-six independent molecular dynamics simulations.
At large volumes, the calculated
excess pressure coincides with the first correction to the
ideal gas law which can be analytically calculated from the 
first virial coefficient of the binary mixture, $B_2(T)$, 
which in the case of our
system is equal to $B_2(T=5.0)=0.53622$. 
To decrease the numerical integration
error we analytically calculate the contribution to the integral
arising from the first virial corrections and integrate over the
volume only $ P_{ex} -B_2(T) k_BT (N/V)^2$. 
As a result, we obtain $S(C)/k_B=8061.7$\cite{details}

\begin{figure}
\centerline{\psfig{figure=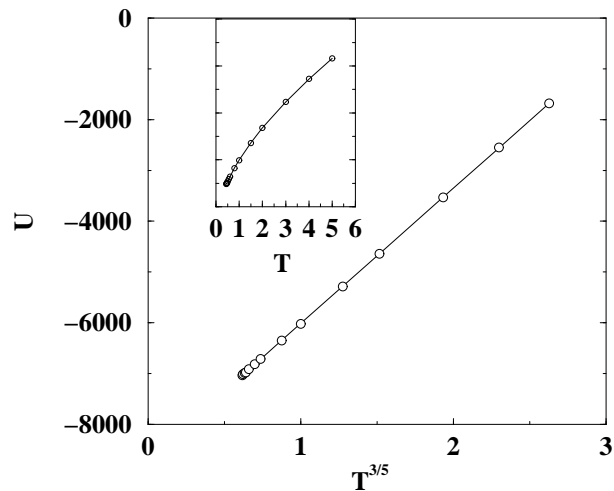,width=8.0cm,angle=0}}
\caption{Potential energy as a function of $T^{3/5}$ for the BMLJ system.
The inset show the same data in liner scale. The continuous line 
is the fit to $U=a+b T^{3/5}$ 
}
\label{fig:talla35}
\end{figure}

The entropy at any $T$ along the studied isochoric path 
can then be calculated 
as
\begin{equation}
S(T,\rho=1.2)= S(C) +
\int_{T=5.0}^{T} {C_V(T') \over T'}  dT' 
\end{equation}
where $C_V(T) = dU(T)/dT+ 3/2N k_B  $ is 
calculated from the $T$-dependence of the system
average potential energy  $U$ obtained from the simulations.
We find that, in agreement with recent theoretical predictions\cite{talla35}, 
the $T$ dependence of $U$  along the studied 
isochore is very well
described by the law $U(T) \sim  T^{3/5}$ (See Fig.\ref{fig:talla35}),
which produces a contribution to the liquid entropy varying as 
$ T^{-2/5}$.
The use of the $ T^{-2/5}$ law provides a reliable extrapolation of
$S_{liquid}$ below the lowest studied $T$.

\begin{figure}
\centerline{\psfig{figure=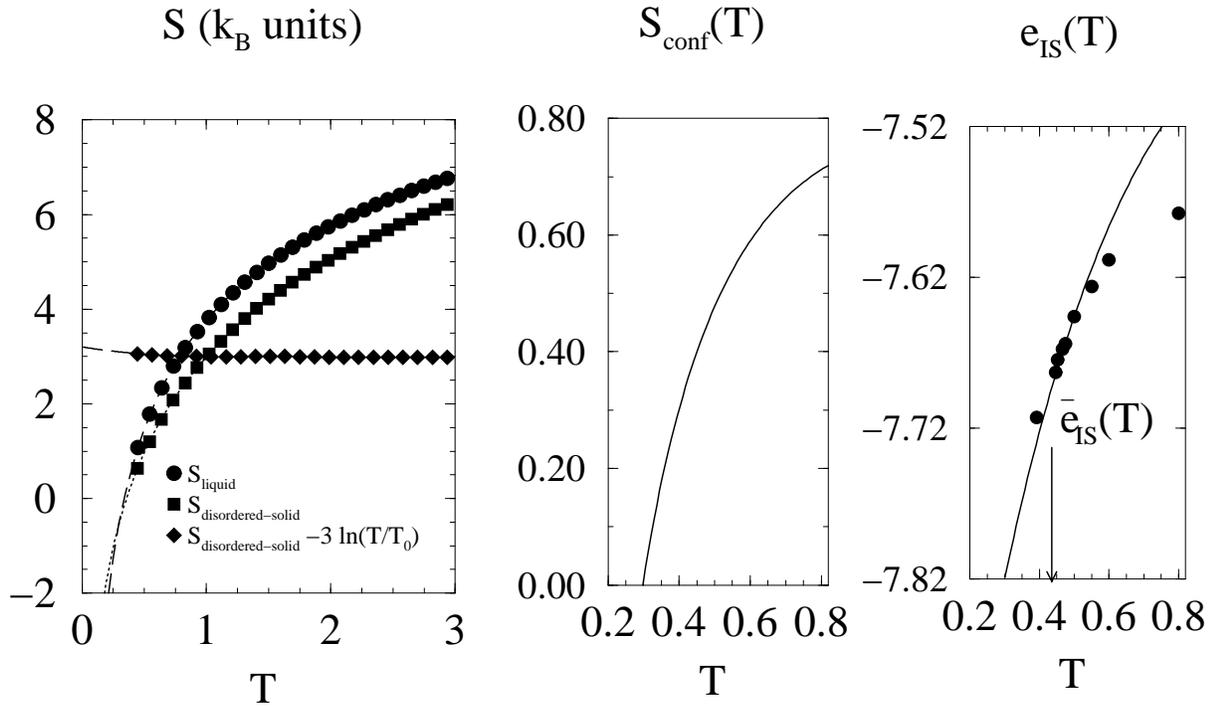,width=16.0cm,angle=0}}
\caption{Left: Liquid (circles) and disordered-solid (squares) 
entropies as a function of $T$. The diamonds show the $T$ dependence of 
the disordered solid entropy, once the explicit $T$ dependence is 
subtracted, to highlight the weak residual $T$ dependence due to the
$T$-dependence of the density of states. Such weak residual $T$-dependence
has been extrapolated to lower $T$ via a quadratic fit and used to
provide an analytic expression for the $T$ dependence of the
disordered solid entropy. Middle: $T$-dependence of the configurational entropy. 
Right:   $T$-dependence of the IS energy
for the BMLJ system as determined from the simulation (circles) and from
Eq.~(\protect\ref{eq:sdie}) (solid line).}
\label{fig:entropy}
\end{figure}

To estimate the basin entropy, 
we 
assume that at the lowest studied $T$, the unknown  
$f(\beta,e)$ can be approximated by the harmonic free energy of a
disordered system characterized by the eigenfrequencies spectrum
calculated from the IS at
the corresponding $T$. In this approximation, 
the difference between the entropy of the liquid and the
entropy of the harmonic disordered solid coincides with the
configurational entropy.
We evaluate the entropy of the disordered solid in the 
harmonic approximation as

\begin{equation}
S_{disordered-solid}(T,V)= \sum_{j=1}^{3N-3} 1 - \ln (\beta  \hbar \omega_j)
\end{equation}

\noindent
where $\omega_j$ is the frequency of the 
$j$-th normal mode. The $e_{IS}$-dependence in  
$S_{disordered-solid}(T,V)$ enters via the $e_{IS}$-dependence of the
density of states. Consistently with the estimate of the
$e_{IS}$ dependence of $f(e_{IS},T)$ discussed below, we find (see Fig.
\ref{fig:entropy}-left)
that the $T$ dependence of the density of states accounts for only a few percent of
the $T$-change in $S_{disordered-solid}$ at low $T$.

The $T$ dependence of the evaluated liquid and disordered solid entropies
is reported in Fig.\ref{fig:entropy}-left. The $T$
dependence of the configurational entropy (difference between
$S_{liquid}$ and $S_{disordered-solid}$) is reported in 
\ref{fig:entropy}-middle.
We note that, if the extrapolations are reliable,
the configurational entropy
vanishes at $T=0.297 \pm 0.01 $, which defines the Kauzmann temperature 
$T_K$\cite{kauzman}
for the studied binary mixture, in agreement with the
findings of Coluzzi et al\cite{btesi,barbara}.
Note that the configuration entropy 
around  $T_{MCT}=0.435$ is halfway between $T_K$  and the
high $T$ value, suggesting that the ordering process in configuration space
at the lowest 
temperature which we have been able to equilibrate is far 
from being complete. Of course, the present data do not furnish a full 
proof of 
the existence of a finite $T$ at which $S_{conf}$ goes to zero, being based 
on a large (but apparently reliable, see Fig.\ref{fig:talla35}) 
extrapolation in $T$. 
We also note that the ratio between $T_K$ 
and $T_{MCT}$  support the view that the studied system has
intermediate fragility character, as recently predicted by Angell and
coworkers on the basis of a comparison between 
experimental results and numerical data for the
same system\cite{austen-pisa}.

\section{IS-energy dependence of the configurational entropy}
\label{sec:entropyE}

In this section we show that in the BMLJ case, for $T< 0.8$, the
factorization approximation discussed in 
Sec.~\ref{sec:theorylowT} is indeed satisfied. The possibility of
separating the IS subsystem thermodynamics from the 
basin thermodynamics allow us to calculate the 
$e_{IS}$ dependence of the configurational entropy and thus to estimate
the number of basins in configuration space with the same $e_{IS}$ energy.

To test the validity of the factorization approximation, we
evaluate the lhs of Eq. \ref{eq:pofe}, i.e. we calculate 
the $e_{IS}$ dependence of  $\ln(P(e_{IS},T))+e_{IS}/T$.
As discussed in Sec.~\ref{sec:theorylowT}, if  
$f(\beta,e_{IS})$  has only a weak dependence on  $e_{IS}$, then 
it must be possible to 
superimpose curves at different temperatures
which overlap in $e_{IS}$.
Then, the resulting $e_{IS}$-dependent curve
is, except for an unknown constant, $S_{conf}(e_{IS})$,
in the $e_{IS}$ range accessed at the studied $T$.

\begin{figure}
\centerline{\psfig{figure=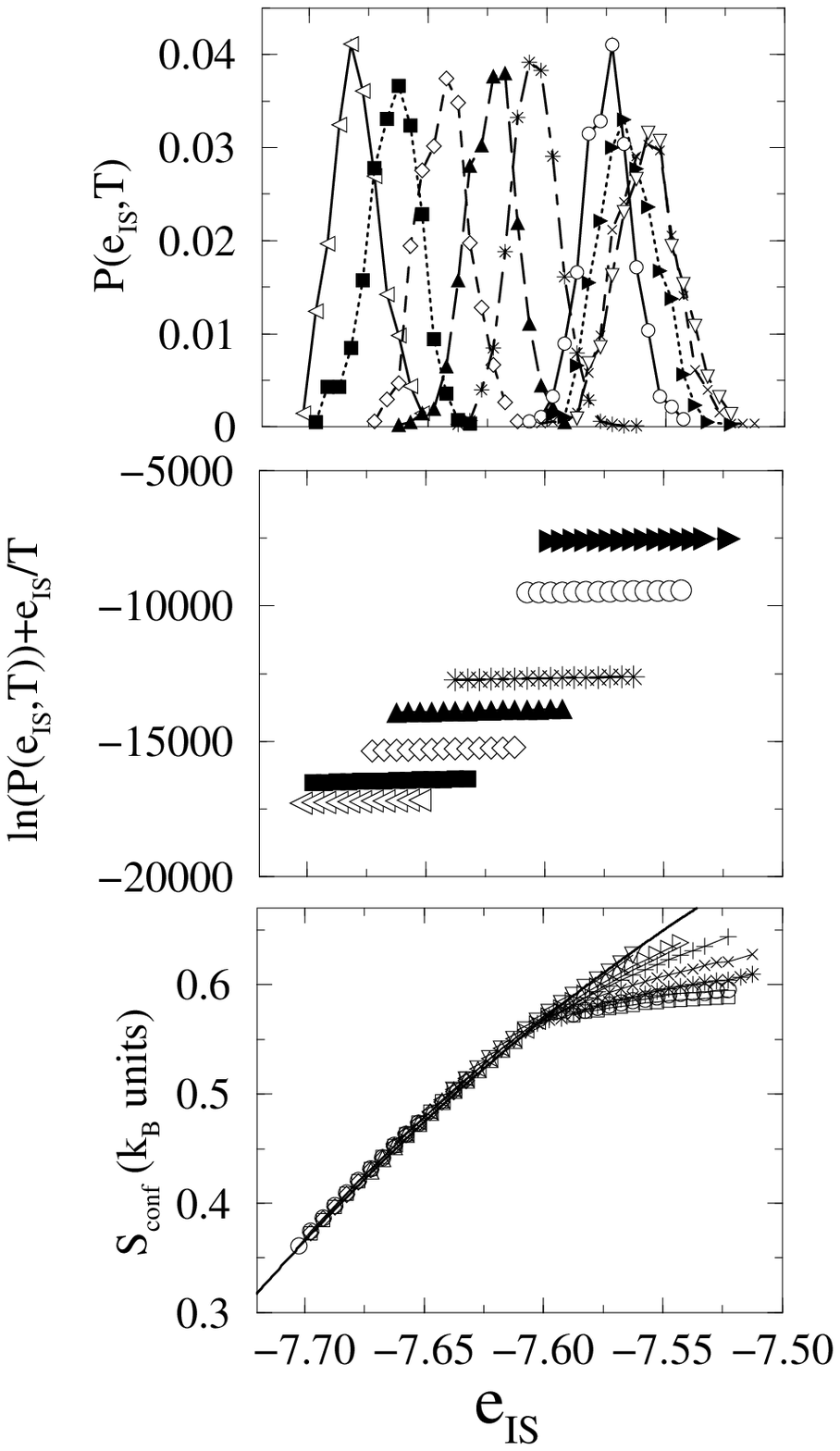,width=8.0cm,angle=0}}
\caption{Top: Distributions $ P(e_{IS},T)$  of the $IS$ energy (per atom)
for different equilibrium temperatures $T$.  ~From left to right:
$T=0.446, 0.466, 0.5, 0.55, 0.6, 0.8, 1.0, 2.0, 4.0$.
Middle:  $\ln[P(e_{IS},T)]+ \beta e_{IS} $,
for six different equilibrium temperatures $T$.
Same symbols as in the top panel.
Bottom: Data in the middle panel (plus data corresponding to other $T$) 
are displayed shifted to maximize the
overlap between
curves with different $T$ {\it and} the overlap with
$S_{conf}(e_{IS})$ (in absolute units), calculated as discussed in the text and shown here  as full line.
The curves which do not lay on the continuous line correspond to
$T=5.0,4.0,2.0,1.5,1.0,0.8$, from bottom to top.
}
\label{fig:pofe}
\end{figure}

This procedure is displayed in Fig.\ref{fig:pofe}.
We note that while
below $T=0.8$ curves for different $T$ lie on the same master curve, 
above $T=0.8$, curves for different $T$ have different $e_{IS}$
dependence, highlighting the progressive $e_{IS}$-dependence of
$f(\beta,e_{IS})$.

The data presented in Fig. \ref{fig:pofe} 
are particularly relevant. They show that, below $T=0.8$, 
the IS can be treated as a system of levels characterized by
an energy value $e_{IS}$ and an associated degeneracy $\Omega(e_{IS}) $.
Thus, for the $e_{IS}$ subsystem
it is possible to use the standard thermodynamics relations
to evaluate the $T$-dependence of the average energy and entropy. In this
respect,  the $T$ dependent 
configurational entropy (but only below $T=0.8$) 
can be evaluated as 

\begin{equation}
{{ d {S}_{conf}(T)} \over {d {e}_{IS}(T)}}=   {{1}\over{T}} ~~~~~~~~~~
e_{IS}(T) = e_{IS}(T_K)+\int_{T_k}^T  T  d S_{conf}(T)   
\label{eq:sdie}
\end{equation}
\noindent
By integrating the configurational entropy from $T_K$ upward,
it is possible to calculate the
the $T$ dependence of $e_{IS}$. 
The unknown integration constant $e_{IS}(T_K)$
can be calculated by comparing the obtained expression 
with the $e_{IS}(T)$-dependence calculated directly from the
simulation in the region $T \le 0.8$ (see also Ref.\cite{nature})
The present analysis (see Fig. \ref{fig:entropy}-right) predicts  $e_{IS}(T_K)= -7.82 \pm 0.01$.

~From $S(T)$, evaluated in the previous Section, and
from $e_{IS}(T)$, evaluated according to Eq. (\ref{eq:sdie}),
it is possible to eliminate the $T$ dependence and to calculate the
 $e_{IS}$ dependence of the configurational entropy in an absolute 
scale, which can be compared with the one calculated independently ---
but with an unknown constant --- via the superposition of the different
$\ln(P(e_{IS},T))+e_{IS}/T$ curves.
Such a comparison is done in Fig. \ref{fig:pofe}.
The agreement between the two set of measurements confirms the
validity of the analysis presented in this article and the
quality of the factorization approximation.

Before concluding this section, we note that
an estimate of the $e_{IS}$ dependence of the configurational
entropy, based on the analysis of experimental data have been presented
in Ref.\cite{stillingerotp}. Analysis of the configurational entropy 
as a function of internal system parameter's (which conceptually are 
equivalent to the $e_{IS}$ choice adopted in the present work) have been reported in Ref. \cite{speedyperaustin,speedydeben}

\section{Conclusions}

The data and the analysis reported in this article offer a detailed
thermodynamic description of the supercooling state, based on the
formalism proposed by Stillinger and Weber.
In particular we have presented a
quantitative evaluation of the 
degeneracy of the inherent structures  (which before was only 
calculated for systems composed by less than 50 atoms\cite{stillingersearxh,
heuer}) for a bulk system. We consider particularly relevant
the presented evidence that 
in the supercooled states  (below $T=0.8$ for the studied system) 
the thermodynamics of the  inherent structures  almost completely decouples
from the ``vibrational'' thermodynamics (i.e. from the process of 
exploration of the IS basin).
It is particularly important to notice that 
a thermodynamics
approach for the inherent structures subsystem becomes 
possible in supercooled states.  
The description of supercooled liquids as composed by two weakly coupled
subsystems --- the IS subsystem and the ``vibrational'' basin subsystem ---
offers stimulating ideas both for a microscopic understanding of the
out-of-equilibrium thermodynamics recently proposed\cite{teo}
(since if the factorization were exact, one could think of keeping the
two subsystem coupled to two different temperatures) 
and the aging processes\cite{cugliandolo}, as well as 
for the still missing theoretical quantitative
description of the slow dynamics below the MCT temperature.
A first step in the direction of estimating the temperature at which
the configurational subsystem is 
in quasi-equilibrium during an aging process 
has been recently reported \cite{aging,kobthisvolume}.

Finally, we stress that
the description we have presented refers to a constant volume 
system. In this respect, it is based on one internal parameter only
(in the language of Davies and Jones\cite{davies}), 
which we have identified with 
$e_{IS}$. In a full treatment, at least one other internal
parameter would be necessary, to discriminate between basins with the
same $e_{IS}$ but different volume.
We plan to further test the validity of such one-internal
parameter description for isochoric cooling.

\section{Acknowledgments}

We thank G. Parisi for bringing to our
attention Ref.\cite{talla35} and for suggesting us to extrapolate the
$T$ dependence of the entropy according to $S \sim T^{-2/5}$. 
We thank A. Scala, E. La Nave, P. Poole
for comments. F.S. and P.T. acknowledge support from MURST PRIN 98 and INFM PRA99
and W. K. from the DFG though SFB 262.
F.S. acknowledges stimulating discussions with R. Speedy.


\begin{thebibliography}{999}
\bibitem{details}
In Eq. (\ref{eq:entro}) we have used 
$\hbar=0.063507 kJ/mol~ps$. We have assumed the AA LJ interaction
equal to $1~kJ/mol$, the unit of length equal to $1~nm$, the unit
of time $1~ps$. With this choice, each the three terms in Eq. (\ref{eq:entro})
contribute $10734.1, 1678.5/5, 2336.7$ respectively.
\end{thebibliography}
\end{document}